\documentclass{epl}
\usepackage{epsfig}
\usepackage{graphicx}

\title{The role of discrete-particle noise in the Ostwald ripening}
\shorttitle{Discrete-particle noise in the Ostwald ripening}
\author{Baruch Meerson\inst{1}, Leonard M. Sander\inst{2} \and Peter Smereka\inst{3}}
\shortauthor{B. Meerson \textit{et al}.}

\institute{\inst{1} Racah Institute of Physics, Hebrew University
of
Jerusalem, Jerusalem 91904, Israel \\
\inst{2} Michigan Center for Theoretical Physics, Department of
Physics, University of Michigan, Ann Arbor, MI 48109-1120, USA\\
\inst{3} Department of Mathematics, University of Michigan, Ann
Arbor, MI 48109-1120, USA}




\begin{document}

\maketitle

\begin{abstract}
We investigate the role of discrete-particle noise in
interface-controlled Ostwald ripening. We introduce the noise
within the framework of the Becker-D\"{o}ring equations, and
employ both Monte Carlo simulations and direct numerical solution
of these equations. We find that the noise drives the system
towards a unique scaling regime describable by a limiting solution
of a classical continuum theory due to Lifshitz, Slyozov and
Wagner. The convergence towards the scaling solution is
\emph{extremely} slow, and we report a systematic deviation
between the observed small correction to scaling and a theoretical
prediction of this quantity.
\end{abstract}

\section{Introduction}

Ostwald ripening (OR) is a generic coarsening process which occurs
in a late stage of phase separation, when the domains (or
clusters) of the minority phase compete for monomers. As a result,
the larger clusters grow at the expense of the smaller ones.
Following the pioneering work of Lifshitz and Slyozov \cite{LS}
and Wagner \cite{W}, experimental and theoretical investigations
of OR have focused on the dynamic scaling properties of the
probability distribution function (PDF) of cluster sizes, and of
its moments. These properties are determined by the kinetics of
the monomer transport. Simple limits of this kinetics are observed
when the monomer transport is controlled either by diffusion of
the monomers in the bulk, or by the processes of attachment and
detachment of the monomers at cluster interfaces. The latter limit
is  called interface-controlled. In each of the two limits
continuum mean-field theories have been formulated: for
diffusion-controlled OR by Lifshitz and Slyozov (LS) \cite{LS},
and for  interface-controlled OR by Wagner \cite{W}. The two
models are often united under the name of the LSW model.  The LSW
model admits a family of self-similar solutions for the PDF of
cluster sizes. However, the problem of selection of the correct
self-similar solution is non-trivial. All of the LSW scaling
functions have compact support and can be parameterized by the
value, $\lambda$, of the logarithmic derivative at the edge of the
support; $-1< \lambda\le \infty$. LS \cite{LS} argued that the
selected PDF for the case of initial conditions with a long tail
is the limiting PDF, corresponding to $\lambda=\infty$. The scaled
PDF should approach this function as $t \to \infty$. On the other
hand,  if the initial data for the PDF has compact support the
selected self-similar PDF is determined by the behavior of the
initial data near the edge of its support \cite{MS,GMS,NP}. If the
initial PDF has  logarithmic derivative $\lambda_0$ at the edge of
support, the selected self-similar solution is the one with the
same value of the logarithmic derivative at the edge of its
support: $\lambda=\lambda_0$ \cite{MS,GMS}. If the logarithmic
derivative of the PDF at the edge of support at $t=0$ does not
exist, the PDF does not approach any self-similar solution
\cite{NP}.

These results are in apparent contradiction to experimental
results which appear to show strong selection, \textit{i.e.}
selection insensitive to  initial conditions. To find strong
selection we must consider dynamics beyond that of the classical
LSW model. One possibility is to account for discreteness of atoms
in any real system. This direction was explored by Vel\'{a}zquez
\cite{V} and Meerson \cite{M}, who employed the Becker-D\"{o}ring
(BD) equations \cite{BD}, properly modified to account for
conservation of the total number of atoms \cite{Binder,Burton}. In
the limit of $\bar{s} \gg 1$, where $\bar{s}$ is the average
number of atoms in a cluster, the BD-equations reduce to the LSW
model \cite{Binder}. Taking into account the next order term in
$1/\bar{s}$,  one arrives at a Fokker-Planck (FP) equation for the
cluster size PDF \cite{V,M}, see below. The drift term of the
FP-equation coincides with that of the LSW model, while the small
diffusion term comes from discrete-particle noise. The diffusion
term produces a tail in the PDF, even if the initial data has
compact support at $t=0$. According to Refs. \cite{V,M} (see also
Ref. \cite{RZ}), this tail drives the system towards the limiting
self-similar solution corresponding to $\lambda =\infty$.

These arguments assume, however, that the FP-equation is a
faithful long-time description of the BD-equations. Though
natural, this assumption is not obviously correct. There are many
examples when the FP equation misses important aspects of discrete
systems \cite{rare}. 
In this work we investigate  the role of discrete-particle noise
by  dealing directly with the BD equations, without making the FP
approximation. We focus on interface-controlled kinetics. This
choice is motivated by experimental findings which showed the
importance of this (sometimes overlooked) limit in a variety of
environments, such as the coarsening of two-dimensional
islands on Si(001) 
\cite{Bartelt}, coarsening of granular clusters in
electrostatically driven granular powders \cite{Aranson}, etc.
There is an important additional motivation. Both the LSW model
and the BD-equations neglect spatial correlations. As a result,
the quantitative validity of the LSW model is limited, in the case
of the diffusion-controlled OR, to extremely small area fractions
\cite{VoorheesMarder}. By contrast, spatial correlations in
interface-controlled OR are much weaker \cite{Conti}. Therefore,
for the same value of area fraction, the BD-equations are more
accurate in describing interface-controlled OR, than
diffusion-controlled OR \cite{merger2}.

\section{BD-equations and the LSW model}

Let $s$ be the number of atoms in a cluster, and $N_s (t)$ be the
number of clusters of size $s$. The BD-equations
\cite{BD,Binder,Burton} are master equations for the populations
of clusters, $s\ge 2$:
\begin{equation}
\label{BD1}\dot{N_s} = N_1 (K_{s-1} N_{s-1} - K_s N_s)
-\frac{N_s}{\tau_s} +\frac{N_{s+1}}{\tau_{s+1}}\,,
\end{equation}\
and monomers:
\begin{equation}
\label{BDmon} \dot{N_1} = -2 K_1 N_1^2 - N_1 \sum_{s \geq 2}K_s
N_s + \frac{2 N_2}{\tau_2}+\sum_{s\geq3}\frac{N_s}{\tau_s}\,.
\end{equation}
Here $K_s=K_1 s^p$ is the rate of attachment of monomers to the
cluster of size $s$, and $\tau_s=a s^q$ is the inverse rate of
detachment of monomers from the cluster of size $s$. One can
always choose $a=1$: this corresponds to a rescaled time $\tilde
{t}=t/a$ (the tilde will be omitted in the following), and
a rescaled attachment rate coefficient $\alpha = K_1 a$. 
Equations (\ref{BD1}) and (\ref{BDmon}) preserve the total
number of atoms $N$:
\begin{equation}
\label{BD2} N_1 +\sum_{s=2}^{s_{max}} s\,N_s = N\,.
\end{equation}

To set the stage for our analysis, let us briefly review the
predictions of the underlying continuum theories. When the
dynamics (\ref{BD1}) and (\ref{BDmon}) reach the stage of OR, one
can proceed to the limit of $s \gg 1$ and treat $s$ as a continuum
variable (except for the monomers, $s=1$, which should be taken
care of separately). Then, by a truncated Taylor expansion, one
obtains the FP-equation \cite{V,M}
\begin{equation}
\frac{\partial n_s}{\partial t} + \frac{\partial}{\partial
s}\left(V_s n_s \right) =\frac{1}{2}\,\frac{\partial^2}{\partial
s^2} \left(D_s n_s \right)\,, \label{FP}
\end{equation}
where
\begin{equation}
V_s (t) = A n_1 s^p  -s^{-q} \quad \mbox{and} \quad D_s (t) = A
n_1 s^p  + s^{-q} \label{coeff}
\end{equation}
are the drift velocity and the diffusion coefficient in
$s$-space, $n_s(t)=N_s(t)/N$ and $A=K_1 a N$.

The LSW model neglects the diffusion term in Eq. (\ref{FP}) and
deals with the continuity equation
\begin{equation}
 \frac{\partial n_s}{\partial t}+ \frac{\partial}{\partial
s} \left[(A n_1 s^p  -s^{-q}) n_s \right] =0\,, \label{LSW}
\end{equation}
combined with the conservation law
\begin{equation}
\label{cons1} n_1 +\int_{0}^{\infty} s\,n_s (t) \,ds = 1\,,
\end{equation}
see Ref. \cite{M} for details. For interface--controlled OR one
obtains $p=1/2$ and $q=0$ in two dimensions, and $p=2/3$ and
$q=-1/3$ in three dimensions \cite{M}. In the following we will
focus on the two-dimensional case. At late times the contribution
of the monomers to the total number of atoms in the system becomes
negligible, and Eq. (\ref{cons1}) becomes
\begin{equation}
\label{cons2} \int_0^{\infty} s\,n_s (t) \,ds \simeq 1\,.
\end{equation}
It follows from Eqs. (\ref{LSW}) and (\ref{cons2}) that
\begin{equation}\label{relation}
 s_c^{1/2}(t)=\langle s^{1/2} \rangle\equiv\frac{\int_0^{\infty}
  s^{1/2} n_s\,ds}{\int_0^{\infty}
 n_s\,ds}\,,
\end{equation}
where $s_c=(A n_1)^{-2}$ is the time-dependent critical size of
the clusters; that is clusters with $s>s_c$ grow, while clusters
with $s<s_c$ shrink. Equations (\ref{LSW}) and (\ref{cons2}) admit
a family of self-similar solutions:
\begin{equation}
\label{ansatz}
 n_s = \frac{1}{s_c^2} \,\phi_{\beta} \left(\frac{s}{s_c}\right), \quad \quad
 n_1=\frac{\beta}{A
 t^{1/2}}\,,
\end{equation}
parameterized by $\beta=\mathrm{const}$. According to Ref. \cite{V,M}, the
discrete-particle noise selects the \textit{limiting} solution
with $\beta=2$. The corresponding scaling function \cite{W,Hillert}
\begin{equation}\label{wagner2d}
\phi(x)=\left\{\begin{array}{ll} \frac{C}{\left(2 -
\sqrt{x}\right)^4} \, \exp \left(-
\frac{4}{2 - \sqrt{x}} \right)\,& \mbox{if $0<x<4$}, \\
0\, & \mbox{if $x\ge 4$}\,,
\end{array}
\right.
\end{equation}
has an infinite number of vanishing derivatives at the edge of its
support $x_{max}=4$ and $\lambda=\infty$. In Eq. (\ref{wagner2d})
$C=e^{-2}+2\, \mathrm{Ei}(-2)=26.6423\dots $, where $\mathrm{Ei}$
is the exponential integral function \cite{Abramowitz}.

\section{Monte-Carlo simulations}

Direct Monte-Carlo simulations of interface-controlled OR are
inefficient, because most of the computation time is spent to
resolve the fast monomer diffusion in the bulk, while the kinetic
bottleneck  here is the slow attachment and detachment of monomers
at the clusters interface \cite{Zhdanov}. To accelerate the
simulations  we assume that the monomer transport in the bulk is
\emph{instantaneous}. Therefore, we deal with a collection of
clusters, $s\geq 2$, and a pool of monomers, without taking care
of the spatial distribution of any of them. Essentially, this
corresponds to a stochastic simulation of the BD equations. We
start with a collection of clusters in some initial condition, and
some initial number of monomers, $N_1$. We repeatedly choose a
cluster at random and let it either grow, by absorbing monomers
with rate $\alpha N_1s^{1/2}$, or shrink by emitting monomers with
unit rate.

This is achieved by rejection Monte Carlo: we calculate the rate
for the largest cluster, $s=s_{max}$, to either grow or shrink,
$f_{max}=\alpha N_1 s_{max}^{1/2} +1$, and the corresponding
quantity for the cluster at hand, $f=\alpha N_1 s^{1/2} +1$. Then
we choose a random number $u$ between 0 and $f_{max}$. If $u>f$ we
do nothing. If $1 < u \le f$ we make the cluster grow by
increasing $s$ by 1, and decreasing $N_1$ by 1. If $u \le 1$ we
shrink the cluster, that is decrease $s$ by 1 and increase $N_1$
by 1. We need to adjoin a special rule for clusters of size 2. If
they shrink we add \emph{two} monomers to $N_1$ and remove the
cluster in question from the list. The time was advanced, at each
step,  by $\delta t = 1/(N_{cl} f_{max})$, where $N_{cl}$ is the
current total number of clusters (excluding monomers).

Typical results of these Monte Carlo simulations are shown in
Figs. 1 and 2. Figure 1a depicts the number density of monomers
$n_1$ versus time, and the theoretical prediction for $n_1$ from
Eq. (\ref{ansatz}) with $\beta=2$. Good agreement is observed. The
average cluster size, which we define as $s_a=\langle
s^{1/2}\rangle^2$ is plotted versus time in Fig. 1b.  Linear
growth is observed as expected. The observed slope $0.268$,
however, is slightly higher than the theoretical value $0.25$. The
scaled PDFs of the cluster sizes, $s_c^2(t)\,n_s (t)$, are plotted
versus $x=s/s_c(t)$ at three different times in Fig. 2. Only times
up to $t=3,200$ are used, because at later times the measured PDFs
become too noisy. The three scaled PDFs in Fig. 2 show a good
collapse, but the scaling function is slightly different from the
theoretical prediction (\ref{wagner2d}). Where does the deviation
come from? The LSW-problem notoriously has  very slow convergence
to the limiting self-similar solution \cite{LS,Voorhees2}.
Therefore, one can attempt to attribute the observed deviation to
very slow convergence, masked by fluctuations.
\begin{figure}
\begin{center}
\begin{tabular}{cc}
\epsfxsize=5.2cm  \epsffile{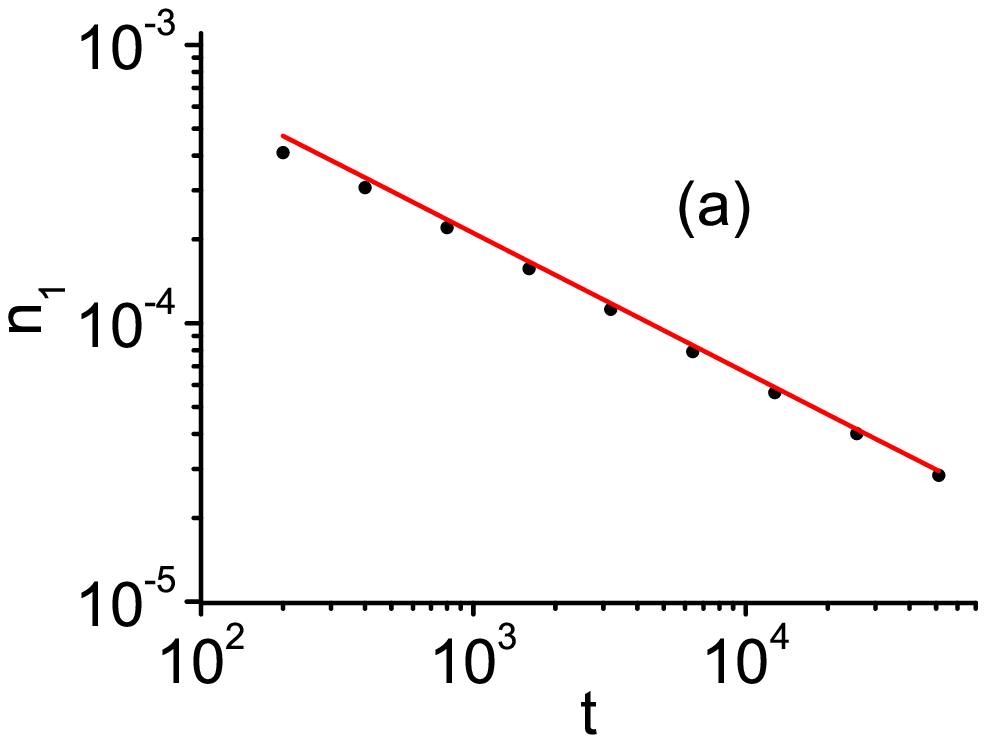} & \epsfxsize=5.2cm
\epsffile{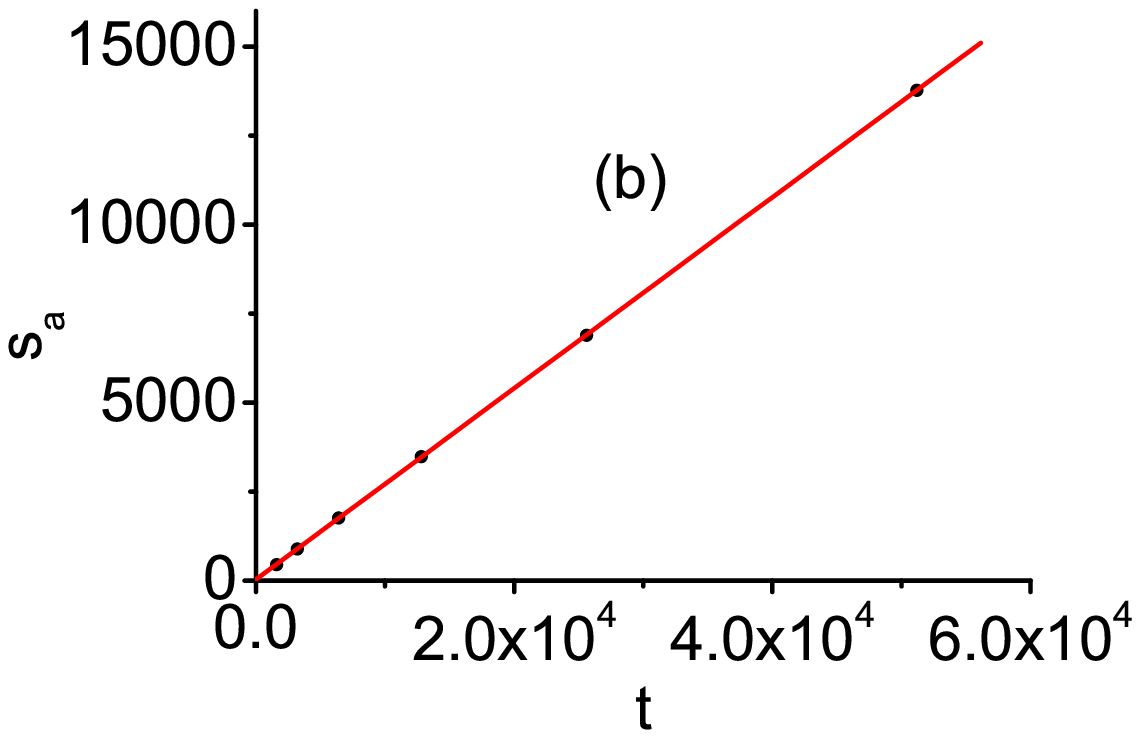}\\
\end{tabular}
\end{center}
\caption{The number density of monomers $n_1$ (a) and the average
cluster size $s_a$ (b) versus time. The circles are averages over
$10^3$ realizations, obtained in Monte Carlo simulations with
$N=3,001 \times 10^6$ atoms. The parameter $\alpha = 10^{-4}$. The
initial conditions are: $N_1=10^3$, $N_{50}=6 \times 10^4$, the
rest of $N_s$ is zero. The red lines show the theoretical
prediction from Eq. (\ref{ansatz}) with $\beta=2$ (a) and a linear
fit of the data (b).} \label{fig1}
\end{figure}

\begin{figure}
\begin{center}
\includegraphics[width=5.7 cm,clip=]{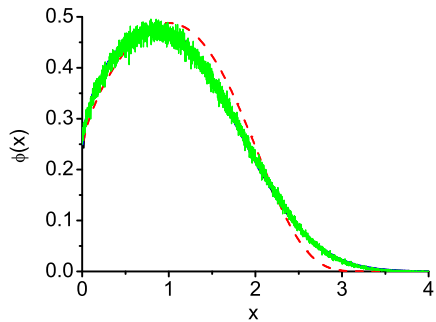}
\end{center}
\caption{The scaled PDFs of the cluster sizes at times $400$
(black), $800$ (blue) and $3200$ (green) as obtained in Monte
Carlo simulations. Because of the relatively large noise at late
times the black and blue lines are masked by the green line. The
parameters are the same is in Fig. 1. The red dashed line is the
theoretical scaling function (\ref{wagner2d}).} \label{fig2}
\end{figure}

\section{Numerical solution of the BD-equations}

To test this interpretation and reach later times, we solved the
BD-equations numerically. Equations (\ref{BD1}) and (\ref{BDmon})
for $1 \le s \le 10^5$ were solved using a fourth order
Runge-Kutta algorithm. The conservation law (\ref{BD2}) was used
for accuracy control. The time step chosen was $\Delta t =  min
\left[ 0.007/(\alpha N_1), 0.125 \right]$. This choice resulted in
excellent mass conservation and enabled us to probe very long
times. For the results reported below only $10^{-7}$ of a single
particle was lost by $t=10^5$.

The results are shown in Figs. 3 and 4. The number density of
monomers $n_1$ versus time (Fig. 3a) agrees very well with the
theoretical prediction.  The average cluster size $s_a=\langle
s^{1/2}\rangle^2$ shows  linear growth with time, see Fig. 3b. The
slope $0.262$, obtained on the interval $10^3<t<10^5$, is still
slightly higher than the theoretical value $0.25$. The scaled PDF
(Fig. 4a), though still different from the theoretical scaling
function (\ref{wagner2d}), apparently slowly approaches it.

\begin{figure}
\begin{center}
\begin{tabular}{cc}
\epsfxsize=5.2cm  \epsffile{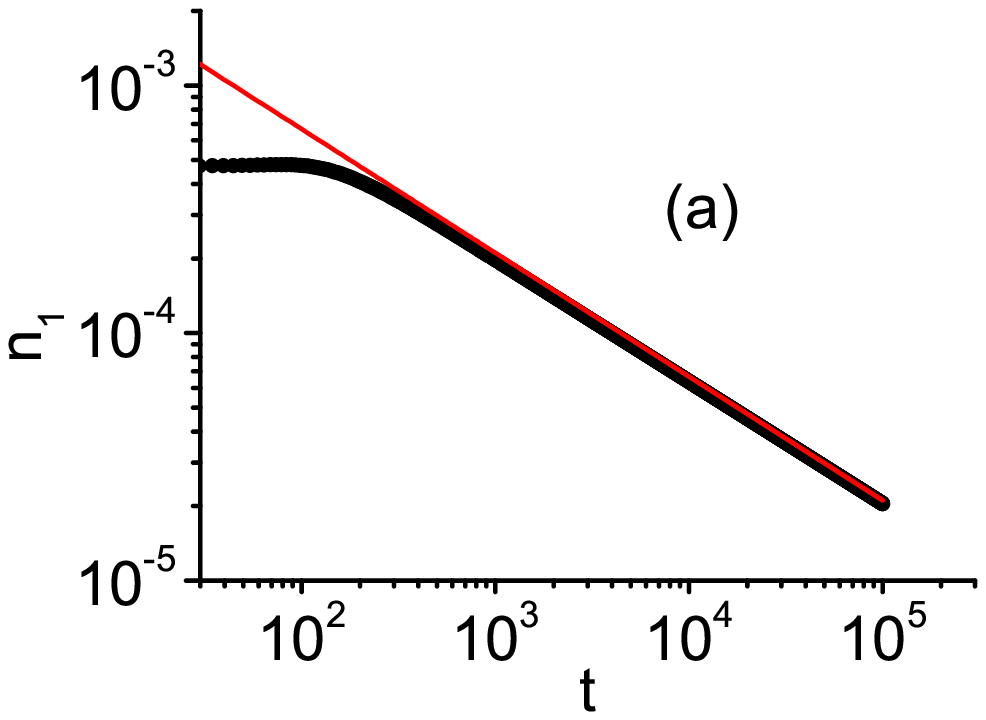} & \epsfxsize=5.2cm
\epsffile{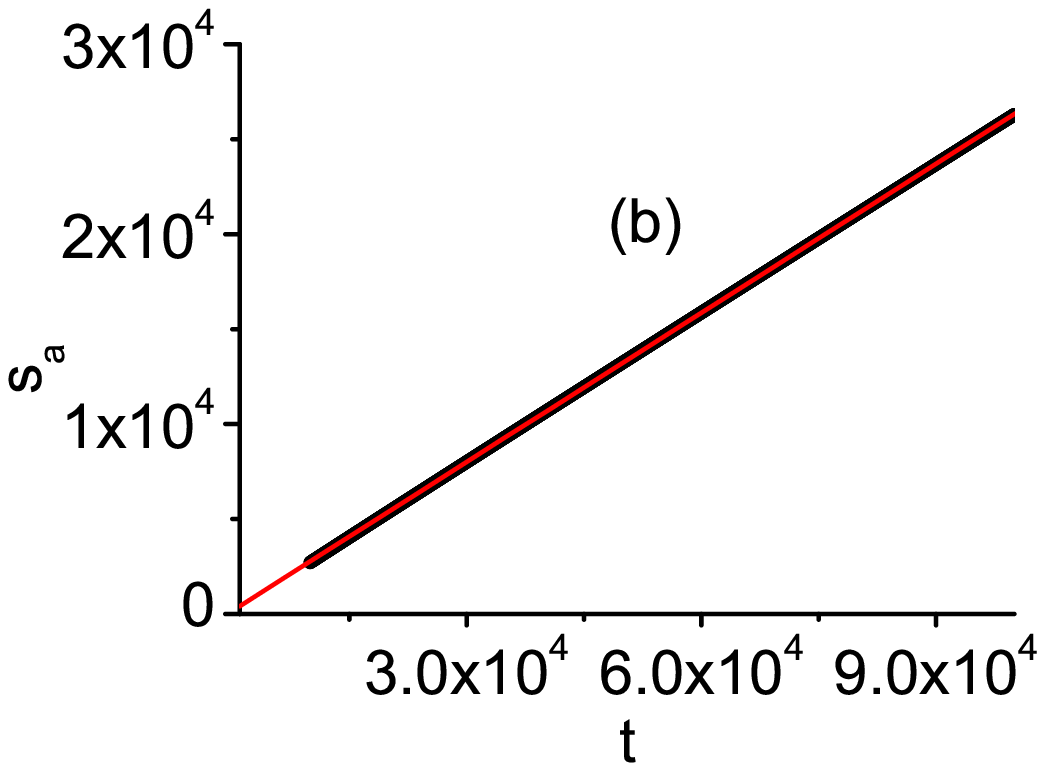}\\
\end{tabular}
\end{center}
\caption{The number density of monomers $n_1$ (a) and the average
cluster size $s_a$ (b) versus time, obtained by solving Eqs.
(\ref{BD1}) and (\ref{BDmon}) numerically for $N=3 \times 10^6$.
The parameter $\alpha = 10^{-4}$. The initial conditions are:
$N_1=1.5 \times 10^3$, $N_{50}=59 970$, the rest of $N_s$ is zero.
The thin red lines show a linear fit of the data (a) and the
theoretical prediction from Eq. (\ref{ansatz}) with $\beta=2$
(b).} \label{fig3}
\end{figure}

\begin{figure}
\begin{center}
\begin{tabular}{cc}
\epsfxsize=5.2cm  \epsffile{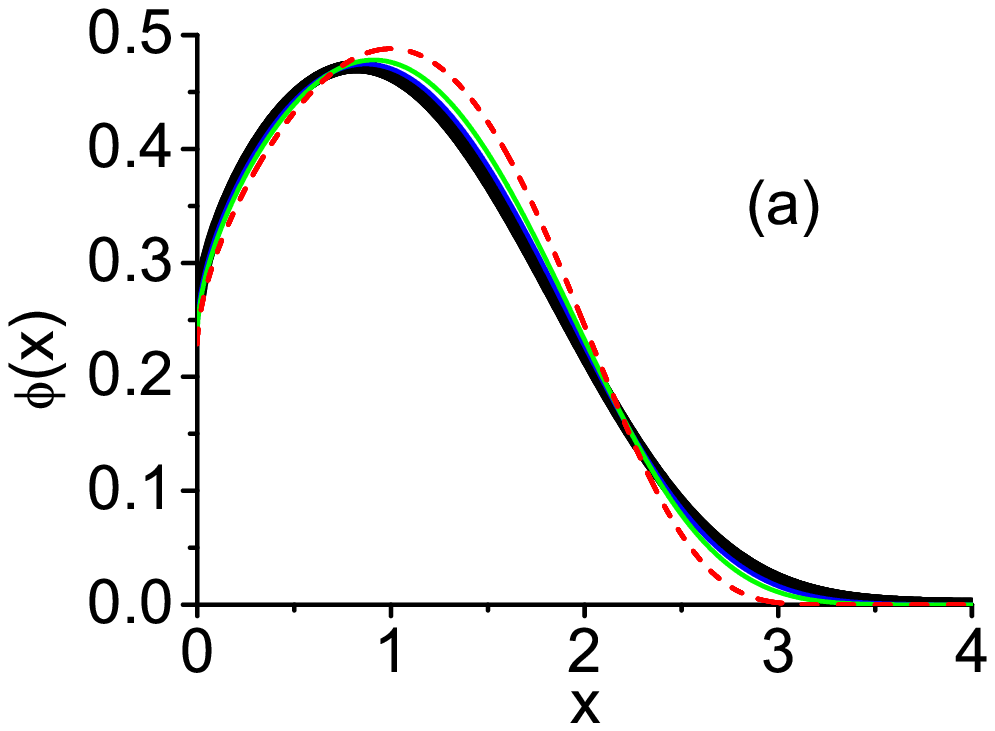} & \epsfxsize=5.2cm
\epsffile{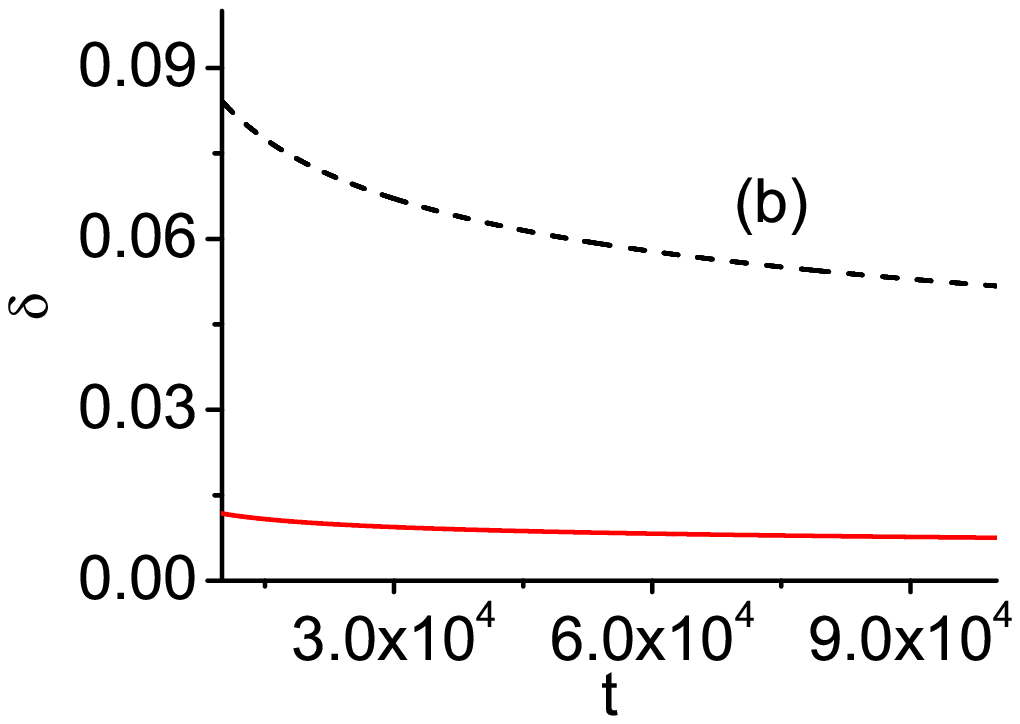}\\
\end{tabular}
\end{center}
\caption{The solid lines in (a) show the scaled PDFs at times $1
350 $ (black), $10 350 $ (blue) and $85 350$ (green), obtained by
solving Eqs. (\ref{BD1}) and (\ref{BDmon}) numerically. The
parameters are the same is in Fig. 3. The red dashed line is the
theoretical scaling function (\ref{wagner2d}). The black dashed
line in (b) shows $\Delta(t)$ (see text), while the red solid line
shows $\ln^{-2} t$, see Eq. (\ref{corr2}).} \label{fig4}
\end{figure}

\section{Logarithmic corrections to scaling}

LS \cite{LS} investigated, in the context of  diffusion-controlled
OR in three dimensions, the convergence towards the limiting
self-similar solution. Here we will employ their argument (see
also Ref. \cite{V}) in the problem of interface-controlled OR in
two dimensions, and compare it with our simulations.

The characteristics of Eq. (\ref{LSW}) are described by the
equation
\begin{equation}\label{char1}
\dot{s} = (s/s_c)^{1/2}-1\,.
\end{equation}
In the rescaled variable $x=s/s_c$ Eq. (\ref{char1}) becomes
\begin{equation}\label{char2}
s_c \dot{x} = -\dot{s_c} x + x^{1/2}-1\,.
\end{equation}
The limiting self-similar solution requires that $s_c(t) \to t/4$
as $t \to \infty$, so that the right hand side of Eq.
(\ref{char2}) becomes a perfect square: $t\, \dot{x} = -
(\sqrt{x}-2)^2$. As shown by LS, this is the only possibility to
have a non-diverging normalization integral (\ref{cons2}), when
the full PDF has a tail.  Following LS, we are looking for the
leading correction in the following form: $s_c(t)=
(t/4)\,[1+\varepsilon^2(t)]$, where $\varepsilon(t) \ll 1$.
Consider a small region of $x$ around the ``blocking point" $x=4$
(the edge of support of the similarity solution). Let
$y(t)=\left[x(t)-4\right]/\varepsilon(t)$. Equations (\ref{char2})
becomes, in the leading order,
\begin{equation}\label{char3}
\frac{1}{4\varepsilon} \frac{dy}{d\tau}=-1+\frac{y}{4}\frac{d}{d
\tau}\left(\frac{1}{\varepsilon}\right) -\frac{y^2}{64}\,,
\end{equation}
where $\tau = \ln t$. By the same normalization argument, the
right hand side should become a perfect square as $t\to\infty$, which
yields $\varepsilon = 1/\tau = 1/\ln t$. Therefore, we obtain
\begin{equation}\label{corr2}
s_a=\langle
s^{1/2}\rangle^2=s_c(t)=\frac{t}{4}\left(1+\frac{1}{\ln^2 t}+
\dots \right)\,.
\end{equation}
The logarithmic correction $\ln^{-2} t$ implies that the
\textit{apparent} dynamic exponent would be slightly larger than
$0.25$ and decreasing in time, as indeed observed in our
simulations. However, a more detailed comparison of Eq.
(\ref{corr2}) with our simulation data is dissapointing. Figure 4b
shows the quantity $\delta = 4 s_c(t)/t-1$ versus $t$. Though
$\delta$ does go down with time (apparently, logarithmically
slowly), it clearly disagrees, at these times, with  the
prediction of Eq. (\ref{corr2}).

\section{Summary}

We have investigated interface-controlled OR in the framework of
conserved BD-equations. We performed Monte-Carlo simulations of
the system, and also solved the BD equations numerically. We
observed that discrete-particle noise drives the system towards a
limiting self-similarity solution described by the LSW theory.
However, the convergence towards the scaling regime is extremely
slow. Furthermore, there is a clear disagreement between the
observed small correction to scaling and a theoretical prediction
of this quantity, obtained in the spirit of the LS theory. At
present we cannot pinpoint the reason for the disagreement in the
subleading order of the theory. Clarifying this important issue
should be the next step of theory.

\acknowledgments

BM is very grateful to the Michigan Center for Theoretical Physics
for hospitality. We acknowledge a useful discussion with Prof.
Juan J.L. Vel\'{a}zquez. BM  was supported by the Israel Science
Foundation (grant No. 180/02). LMS and PS were supported by NSF
grant number DMS 0244419, and  PS was also supported by NSF grant
number DMS 0207402.


\begin{thebibliography}{99}
\bibitem{merger2} Notice that, even
for the interface-controlled OR, the validity of the BD-equations
is limited to a small area fraction, as they disregard cluster
merger. The role of cluster merger in the interface-controlled OR
was investigated in Ref. \cite{Conti}.
\end{thebibliography}
\end{document}